\documentclass[prl,twocolumn]{revtex4}
\usepackage{amsmath}
\usepackage{graphics}

\newcommand \be{\begin{equation}}
\newcommand \ee{\end{equation}}

\begin{document}

\title{\textbf{The Dirac equation and a non-chiral electroweak theory in six dimensional spacetime from a locally gauged $SO(3,3)$ symmetry group}}
\author{C.A. Dartora$^{1}$\thanks{%
cadartora@eletrica.ufpr.br} \ and \ G.G. Cabrera$^{2}$\thanks{%
cabrera@ifi.unicamp.br}}
\affiliation{$^{1}$Electrical Engineering Department, Federal University of Parana
(UFPR), Brazil\\
$^{2}$ Instituto de F\'{\i}sica \ \ `Gleb Wataghin', Universidade Estadual
de Campinas (UNICAMP), C.P. 6165, Campinas 13.083-970 SP, Brazil\\
}

\begin{abstract}
{A toy model for the electroweak interactions(without chirality) is proposed in
a six dimensional spacetime with 3 timelike and 3 spacelike coordinates. 
The spacetime interval $ds^2=dx_\mu dx^\mu$ is left invariant under the symmetry group 
$SO(3,3)$. We obtain the six-dimensional version of the Dirac gamma matrices, $\Gamma _\mu$, and
write down a Dirac-like lagrangian density, ${\cal L}=i \bar{\psi} \Gamma ^\mu \nabla _\mu \psi$. 
The spinor $\psi$ is decomposed into two Dirac spinors, $\psi _1$ and $\psi _2$,
which we interpret as the electron and neutrino fields, respectively. In six-dimensional spacetime
the electron and neutrino fields are then merged in a natural manner. 
The $SO(3,3)$ Lorentz symmetry group must be locally broken to the observable $SO(1,3)$ Lorentz group, 
with only one observable time component, $t_z$. The $t_z$-axis may not be the same at all points of the spacetime and the effect
of breaking the $SO(3,3)$ spacetime symmetry group locally to an $SO(1,3)$ Lorentz group is perceived by
the observers as the existence of the gauge fields. The origin of mass may be attributed to the 
remaining two hidden timelike dimensions. We interpret the origin of mass and
gauge interactions as a consequence of extra time dimensions, without the need of the so-called Higgs mechanism
for the generation of mass. Further, we are able to give a geometric 
meaning to the electromagnetic and non-abelian
gauge symmetries.}


\end{abstract}

\maketitle

The unification of electric and magnetic phenomena into a single theoretical basis
by J.C. Maxwell\cite{[maxwell]} in the 19th century started the search for a 
unified field theory, which presumably will describe all interactions
in nature through a closed set of equations and axioms. The term
{\it unified field theory} was coined by A. Einstein, who unsuccessfully tried 
to merge electromagnetism and gravitation using the geometric language
of general relativity\cite{[einstein]}. Attempts to construct unified field theories follows two 
slightly different approaches. In one hand one has purely geometric theories, which
consider general relativity as fundamental\cite{[kaluza],[rosen],[beil]}. In the other hand there are the 
quantum field theories. It is our opinion that the path followed by Einstein in his hope  
to give a geometrical meaning for the electromagnetic interactions went wrong 
due to the fact that nature must be described by quantum mechanics,  
not taken into account in Einstein's unified field theory. Actually, the usual gravity theory based
on general relativity presents enormous difficulties to be quantized and unified with the other interactions.
Alternative theories have been proposed in order to overcome these problems, as for example
a Yang-Mills gauge theory for gravity\cite{[ghaboussi],[dehnen]}.
The present knowledge of the most successful quantum field theories describing  physics of real 
world is based on two central features:
i) invariance of the physical laws and field equations under the Lorentz-Poincar\'e symmetry group,
namely, the $SO(1,3)$ rotation group and its representations\cite{[Weinberg1]}
and ii) the gauge symmetry principle and symmetry breaking\cite{[Weinberg2],[ryder]}. 
Such theories deal with Minkowskian spacetime having only one timelike coordinate and 
three spacelike coordinates, which are distinguished by the signature of the metric space, 
assumed here to be $(+---)$. In fact, 
there are parallels between gauge theories and 
general relativity, first pointed out by Utiyama already 
in 1956\cite{[uty]}, but the local gauge invariance of the majority of  
quantum field theories is related to {\it internal} symmetries of the fields in an {\it isospace}, 
which in principle is not directly connected to the spacetime geometry. 
From a basic point of view,
the questions of what are these {\it internal} degrees of freedom and why they
exist are not answered yet, to the best of our knowledge. The paradigmatic 
example of a successful quantum field theory is the Weinberg-Salam-Glashow(WSG) electroweak 
theory, which is a non-abelian gauge theory of electromagnetic and weak 
interactions unified into a single theoretical framework\cite{[WSG]}.  
Such theory has two essential ingredients, the first one being the
non-abelian gauge fields, or simply Yang-Mills fields, named after the 1954 seminal 
work by C.N. Yang and R.L. Mills suggesting that isospin 
would be explained in terms of a local gauge theory\cite{[yang]}; the second one is 
the mechanism of spontaneous symmetry breaking, originally proposed by Goldstone and Higgs\cite{[higgs]}, 
allowing to describe the electroweak interactions by means of a gauge field theory. 
Actually, the simplest form of the WSG theory considers two massless Dirac fields describing electrons and
neutrinos which are left invariant under an $SU_L(2)\times U(1)$ local gauge symmetry. We know 
that considerations of gauge invariance and/or relativistic arguments 
require massless gauge fields. However, it is an experimental fact 
that electrons and some of the gauge bosons become massive, at least in low energy regime. 
In order to make the whole theory physically reasonable it is necessary to introduce a scalar field, 
namely the Higgs field, which is responsible for the symmetry breaking of gauge symmetries  in some energy limit through
its interactions with the other fields, attributing masses to some of these other fields. A well succeeded example
of a gauge theory based on the Higgs phenomenon is the Ginzburg-Landau theory of superconductivity, in which
the photon, i.e., the gauge field acquires mass through spontaneous symmetry breaking\cite{[Weinberg2],[ryder]}. 
However, in the WSG electroweak theory the quanta of the Higgs field, the so-called Higgs bosons, were not detected till now. 
In order to overcome this difficulty, it is our aim to obtain a more or less realistic physical theory of 
the gauge and spinor fields interactions without the need of introducing a scalar field. Here
we will appeal to a mechanism which we have called a spacetime symmetry breaking. 
We following the general lines of quantum field theories but we want 
to advance a geometric meaning for the {\it internal} symmetries, which may
be attributed to rotations in hidden spacetime dimensions. 
To go further in the realization of unified field theories,  
a large number of physicists have been considered the hypothesis 
that spacetime has a number of hidden extra dimensions.
Indeed, this is the case of Kaluza-Klein\cite{[kaluza],[klein]}, 
string and supersymmetric theories\cite{[string1],[string2],[string3]}.  
However, the majority of such theoretical models have attributed to these extra dimensions 
a spacelike character. 
In this paper we introduce a six dimensional spacetime structure having three 
timelike and three spacelike coordinates which is left invariant under the ``rotation" group 
$SO(3,3)$. Some attempts in the current literature have been done in order to explore
the Einstein field equations and gravity effects using the $SO(3,3)$ group\cite{[bonelli]}.  
It is clear that time and space components are invariant 
under separate rotations as an $O(3)\times O(3)$ symmetry group embedded
into the larger group.  Actually, in a six-dimensional spacetime space and time are symmetric with
respect to each other, both having the same number of degrees of freedom, but this
larger symmetry must be locally and spontaneously broken to
an $SO(1,3)$ group, which is the local observable spacetime. 
The gauge symmetries will be interpreted as local rotations of the timelike coordinates, leaving
the observable $SO(1,3)$ spacetime invariant at that point. 

As the starting point, consider a general spacetime 
in a given number of dimensions $d=p+q$, being $p$ and $q$ the number
of  timelike and spacelike coordinates, respectively, 
represented by a general $ISO(p,q)$ inhomogeneous Lorentz-Poincar\'e group,
whose spacetime coordinate transformations are given by:
\be
\label{eq1}
x^{\prime \mu} =  \Lambda ^\mu _\nu x^\nu + a^\mu~.
\ee
The matrix $\Lambda ^\mu _\nu$ of the general Lorentz transformations corresponds 
to ``rotations" in this spacetime and the vector $a^\mu$ corresponds to arbitrary translations.
For $a^\mu=0$ we have the homogeneous Lorentz-Poincar\'e group, i.e., the ``rotation" group $SO(p,q)$ 
which is the group of transformations leaving the quadratic norm $ds^2 = dx^\mu dx_\mu$ invariant.
From now on, the four-dimensional spacetime will have coordinates written as
$x^\mu = (t,x,y,z)$, while in six dimensions we assume $x^\mu = (t_x,t_y,t_z,x,y,z)$. 
In quantum mechanics, a given mathematical object transforms according to a specific representation
of the Lorentz-Poincar\'e group.  Aside from the scalar, vector and tensor representations, 
whose transformation properties are defined applying the usual $\Lambda$-matrices, as follows: 
\begin{eqnarray}
U(\Lambda) \phi(x) = \phi(\Lambda x)~,\nonumber\\
U(\Lambda) A^\mu(x) = \Lambda ^\mu _\nu A^\nu(\Lambda x)~,\nonumber\\
U(\Lambda) F^{\mu\nu}(x) = \Lambda ^\mu _\alpha \Lambda ^\mu _\beta F^{\alpha \beta}(\Lambda x)~,\nonumber
\end{eqnarray}
there are mathematical entities, intuitively associated with the quantization
of angular momentum, which transform according to non-trivial 
representations of the Lorentz-Poincar\'e group. They are called {\it spinors}, being essential
for describing almost all known fundamental matter fields.
To obtain a non-trivial representation, i.e., a spinorial representation of the general group, 
it is convenient to follow the Dirac formalism and introduce the anti-commuting gamma matrices. For
a six-dimensional spacetime it can be shown that the minimum dimensionality of these matrices is 
$8\times 8$. In this case these matrices can be constructed from the well known Dirac matrices of the 
four-dimensional spacetime, which obeys the following anti-commuting relation\cite{[Weinberg1],[Weinberg2],[sakurai]}:
\be\label{DiracMat4}
\{\gamma ^\mu,\gamma ^\nu\}=\gamma ^\mu \gamma^\nu + \gamma ^\nu \gamma ^\mu = 2 g^{\mu \nu} {\bf 1}_{4\times 4}~,
\ee
being $\mu,\nu =0,1,2,3$ the spacetime indices and $g^{\mu \nu} = {\rm diag}(+1,-1,-1,-1)$ the metric tensor.
Four of the six gamma matrices required to construct a six-dimensional spacetime algebra, or
Clifford algebra, are given by:
\be
\label{gammamu}
\Gamma ^{\mu} = \left( \begin{array}{cc} \gamma ^\mu & 0\\ 0 & -\gamma^\mu \end{array}\right)=\sigma _z(\gamma ^\mu)~.
\ee
Throughout this paper we will use the notation $\sigma _z(\gamma _\mu)$, which must be
understood as an $8\times 8$ matrix, formally identical to the Pauli matrix $\sigma _z$ but with the numbers
substituted by the four-dimensional $\gamma ^\mu$ Dirac matrices. From now on the temporal index $\mu=0$ of 
the four-dimensional spacetime will be represented by the index $\mu=03$.
The missing $\Gamma$-matrices may be easily obtained observing that
the usual Pauli matrices obey an anti-commuting algebra, $\{\sigma _i,\sigma _j\} = 2 \delta _{ij}$, being $\delta _{ij}$ the
Kronecker delta function. The $8\times 8$ version of the $\sigma _x$ and $\sigma _y$ Pauli matrices,
which anti-commute with the four $\sigma _z(\gamma ^\mu)$, are written explicitly below:
\begin{eqnarray}
\Gamma ^{01} = \left( \begin{array}{cc} 0 & {\bf 1}_{4\times 4}\\ {\bf 1}_{4\times 4} & 0 \end{array}\right) = \sigma _x({\bf 1}), \label{gamamu01}\\
\Gamma ^{02} = \left( \begin{array}{cc} 0 & -i {\bf 1}_{4\times 4}\\ i {\bf 1}_{4\times 4} & 0 \end{array}\right)=\sigma _y({\bf 1}).\label{gamamu02}
\end{eqnarray}
It is straightforward to show that the required six-dimensional spacetime anti-commuting algebra,
\be
\label{DiracMat6}
\{\Gamma ^{\mu},\Gamma ^\nu\} = 2 g^{\mu \nu} {\bf 1}_{8\times 8}~,
\ee
is satisfied by the $\Gamma ^\mu$ matrices,
with spacetime indices given by $\mu = (01,02,03,1,2,3)$ and metric tensor 
$g^{\mu \nu} ={\rm diag}(+1,+1,+1,-1,-1,-1)$.
Now we are in a position to define a Dirac ``bi-spinor", which is, 
in our notation, a set of two 4-component Dirac spinors, $\psi _1$ and $\psi _2$, 
arranged as follows:
\be
\label{DSpinor}
\psi =\left( \begin{array}{c} \psi _1 \\ \psi _2\end{array}\right)~.
\ee 
Following the usual convention we can define the adjoint spinor $\bar{\psi}$:
\[\bar{\psi} =  \psi ^\dagger \Gamma _0 = (\bar{\psi _1},-\bar{\psi _2})~~,\]
where $\bar{\psi _1} =  \psi _1^\dagger \gamma ^0$ and $\bar{\psi _2} =  \psi _2^\dagger \gamma ^0$,
allowing us to write the usual Dirac-like lagrangian density ${\cal L}$:
\be
\label{DiracLagDensity}
{\cal L}=i \bar{\psi} \Gamma ^\mu \nabla _\mu \psi~.
\ee
The six-dimensional derivative  is given by  
$\nabla _\mu = (\nabla _t,\nabla _x)$, but this differential operator
can be broken into the well known  
four-dimensional derivative operator, defined as $\partial _\mu = (\partial _{03}, \partial _i)$, 
being $x_{03}=t$ the observable time component and $\nabla _{t_\perp} = (\partial _{01},\partial _{02}) $ .
The above equation takes a particularly interesting form when written
explicitly in terms of the Dirac spinors $\psi _1$ and $\psi _2$:
\begin{eqnarray}
{\cal L}=i \bar{\psi _1} \gamma^\mu \partial _\mu \psi _1 + i \bar{\psi _2} \gamma^\mu \partial _\mu \psi _2 + \nonumber\\
i \bar{\psi} _1 ( \partial _{01} - i \partial _{02} ) \psi _2- i \bar{\psi} _2 
( \partial _{01} + i \partial _{02} ) \psi _1 ~.\label{LagDensity1}
\end{eqnarray}
Looking at (\ref{LagDensity1}), it is our intention to identify the fields $\psi _1$ and $\psi _2$ 
with the electron and neutrino fields.  For the sake of convenience let us
associate $\psi _1$ with the electron field and $\psi _2$ with the neutrino field.
Despite the experimental evidences showing that neutrinos are massive\cite{[neutrinomass]}, 
we can simplify the picture by considering that the electron field has a mass $m \neq 0$ and the
neutrino is a massless field. In such case we impose the conditions below:
\begin{eqnarray}
(\partial _{01} - i \partial _{02} ) \psi _2 & = & i m \psi _1~,\label{eqmass1}\\
(\partial _{01} + i \partial _{02} ) \psi _1 & = & 0~,\label{eqmass2}
\end{eqnarray}
in order to obtain Dirac equations for a massive electron field 
together with a massless neutrino field. Combining (\ref{eqmass1}) and
(\ref{eqmass2}) it is easy to find a two dimensional Laplace equation in the space of the 
extra time coordinates $(x_{01},x_{02})$:
\[\nabla _{t _\perp}^2 \psi _2 = ( \partial _{01}^2+ \partial _{02}^2 ) \psi _2 = 0~,\]
which can be solved once the boundary conditions are given. 
It is a well known fact that massless fields may become massive when subjected to boundary conditions. 
A trivial example is the electromagnetic field inside a metallic waveguide, in which the massless photon field 
is subjected to the boundary conditions resulting in a energy-momentum dispersion relation similar to that of 
a relativistic massive particle, i.e., the photons inside a waveguide behave as if they are massive\cite{[photonmass]}.
It is tempting to conclude that the mass of the particles is an effect of boundary conditions
imposed on hidden extra dimensions. String theorists proposed long ago  that we can interpret
particles with different masses as modes of vibrations with different boundary conditions 
imposed on a single element, the fundamental string.  As a matter of fact, by making a
slightly modification in equations (\ref{eqmass1}) and (\ref{eqmass2}) we could 
attribute mass to the neutrino field as well, a problem that is not simply soluble 
in the Weinberg-Salam Electroweak theory.

Till now we have shown that in six-dimensional spacetime the electron and neutrino
fields can be represented by same mathematical entity, i.e., they are parts of a single
the Dirac ``bi-spinor" $\psi$. However we are left with the question of how we can
introduce electromagnetic and weak interactions in electroweak interactions in such theory.
Our concern is to show the possibility to obtain a `realistic' theory in which
only the electronic part of the entity $\psi$ interacts with the photon field $A_\mu$ while
the electron and neutrino components of the spinor $\psi$ will interact
with the other gauge fields, namely,  the $W_\mu$ and $Z_\mu$ gauge fields.
To go further, let us obtain the generators of rotation of the time 
coordinates. For an infinitesimal transformation of the form:
\be
\label{inftransf}
\psi^\prime =  \psi + \frac{i}{2} \omega _{\mu \nu} J^{\mu \nu}~,
\ee
being $\omega _{\mu \nu} = - \omega _{\nu \mu}$ an anti-symmetric tensor, it is well known that the
the generators of such transformations will be given by\cite{[Weinberg1],[sakurai]}:
\be
\label{generators}
J^{\mu \nu} = \frac{i}{4}[\Gamma^\mu,\Gamma ^\nu].
\ee
In our six dimensional spacetime, the generators representing rotations of the time components 
are given by the eight-dimensional matrices below:
\begin{eqnarray}
J^{01,02}=J^{03} =\frac{1}{2}\left( \begin{array}{cc} 1 & 0 \\ 0 & -1\end{array}\right) =  \frac{1}{2}\sigma _z({\bf 1})~, \label{j03}\\
J^{02,03}=J^{01} =\frac{1}{2}\left( \begin{array}{cc} 0 & \gamma ^0 \\ \gamma^0 & 0\end{array}\right) = \frac{1}{2} \sigma _x(\gamma ^0)~, \label{j01}\\
J^{03,01}=J^{02} =\frac{1}{2}\left( \begin{array}{cc} 0 & -i \gamma^0 \\ i \gamma ^0 & 0\end{array}\right) = \frac{1}{2} \sigma _y(\gamma^0)~. \label{j02}
\end{eqnarray}
The reader can easily verify that the generators $J^{0i}$ obey an angular momentum algebra,
\be
\label{angmoment}
[J^{0i},J^{0j}]=i \varepsilon _{ijk} J^{0k}~.
\ee  
The effect of pure timelike rotations on the spinor $\psi$ is obtained
applying to it the unitary matrix $U$ defined below:
\be
\label{unitmatrix}
U=\exp\left[i\frac{\vec{\sigma}\cdot{\hat n} \theta}{2}\right] =  
\cos\left(\frac{\theta}{2}\right) {\bf 1}+i \vec{\sigma}\cdot{\hat n} \sin\left(\frac{\theta}{2}\right)~, 
\ee
where the Pauli spin matrices must be understood as 
$\vec{\sigma} = 2 (J^{01},J^{02},J^{03})=[\sigma _x(\gamma^0),\sigma _y(\gamma^0),\sigma _z({\bf 1})]$ here.
An infinitesimal time coordinate rotation implies the following $\psi$-spinor transformation: 
\[\psi^\prime \approx \psi +i \vec{\sigma}\cdot{\hat n} \frac{\theta}{2} \psi~,\] 
which is conveniently written in terms of the Dirac 4-component spinors:
\begin{eqnarray} 
\psi _1^\prime = \psi _1 + i n_z \frac{\theta}{2} \psi _1 +i (n_x-i n_y) \frac{\theta}{2} \gamma ^0 \psi _2 ~,\label{transpsi1} \\
\psi _2^\prime = \psi _2 - i n_z \frac{\theta}{2} \psi _2 +i (n_x+i n_y) \frac{\theta}{2} \gamma ^0 \psi _1 ~.\label{transpsi2}
\end{eqnarray}
Clearly, the lagrangian density (\ref{LagDensity1}) is left invariant under a global rotation of the time coordinates, i.e., 
under the same time rotation at all points of the six-dimensional spacetime.
Let us admit that the Lagrangian is also invariant under an overall phase factor $\chi$, which would 
correspond to a general temporal translation of the origin of the hidden time 
coordinates. We then have the following general transformation on the spinor $\psi$:
\[\psi^\prime  =  \exp\left[i \chi  + i\frac{\vec{\sigma}\cdot{\hat n} \theta}{2}\right] \psi~,\]
corresponding to a gauge symmetry group $SU(2)\times U(1)$.
Now, we want to gauge the above transformation, i.e., we will make $\Lambda$ and $\vec{\theta} =  \theta {\hat{n}}$ functions
of the spacetime coordinates, $x^\mu$. We interpret the local gauge transformation as follows: there is a local freedom 
in the choice of the  $x_{03}=t$ time component, i.e., only the component $x_{03}= t_z$ 
of the time coordinates  $(x_{01},x_{02},x_{03}) = (t_x,t_y,t_z)$ is actually observable, breaking
the $SO(3,3)$ spacetime group to a local $SO(1,3)$ Lorentz group. However, the $t_z$-axis corresponding to the 
observable time coordinate may not be the same at all points of the six-dimensional spacetime, while
the physics is described locally by the Lorentz group $SO(1,3)$ with a single time parameter. The effect
of breaking the $SO(3,3)$ spacetime symmetry group locally to an $SO(1,3)$ Lorentz group is perceived by
the observers as the existence of the gauge fields. In order to keep the lagrangian density invariant
under the local gauge transformations, namely, local rotations and translations in the time components, 
we must introduce gauge fields and replace the ordinary derivatives by their covariant form\cite{[ryder]}:
\be
\label{covderiv}
D _\mu  = \nabla _\mu + i g X_\mu +i g^\prime \vec{\sigma}\cdot{\bf W}_\mu ~.
\ee
The non-abelian or Yang-Mills gauge field $\vec{\sigma}\cdot{\bf W}_\mu$ is introduced
to compensate the effect of local rotation of the time coordinates on the spinor
$\psi$. In the WSG electroweak theory, a similar gauge field is associated to an {\it internal}
space subjected to an $SU_L(2)$ local gauge invariance of the left-handed electron-neutrino 
doublet\cite{[ryder]}, but in our toy model there is no {\it internal} space at all 
and a geometric meaning is possible for the $SU(2)$ gauge sector of the theory, i.e.,
the $SU(2)$ local gauge invariance in our theory is directly associated with rotation
of the time coordinates. Putting (\ref{covderiv}) in place of $\nabla _\mu$ in the
equation (\ref{DiracLagDensity}) we obtain:
\be
\label{LagDensity2}
{\cal L}=i \bar{\psi} \Gamma ^\mu D _\mu \psi = 
i \bar{\psi} \Gamma ^\mu (\nabla _\mu + i g X_\mu +i g^\prime \vec{\sigma}\cdot{\bf W}_\mu)  \psi~.
\ee
Rewriting the interaction terms between the $\psi$ field and the gauge fields in the 
above lagrangian density in the language of the 4-component Dirac spinors, $\psi _1$ and $\psi _2$
lead us to the below expression:
\be
\label{LagDensity2a}
{\cal L}_{int} =  (\bar{\psi _1}~~-\bar{\psi _2} )  \gamma^{\mu}M_\mu \left(\begin{array}{c} \psi _1 \\ \psi _2\end{array} \right)~,
\ee
with the matrix $M_\mu$ defined as:
\be
\label{matrixM}
M_\mu =\left(\begin{array}{cc} - (g X_\mu +  g^\prime W_\mu ^z) & 
- g^\prime (W_\mu^x-i W_\mu ^y) \gamma ^0\\  g^\prime (W_\mu^x+i W_\mu ^y) \gamma ^0 &   
(g X_\mu - g^\prime W_\mu ^z )\end{array}\right) ~.
\ee
In order to achieve our goal, that is, to obtain a toy model for a gauge theory
of electroweak interactions, we follow closely the steps of Weinberg and Salam
\cite{[WSG],[Weinberg2],[ryder]} and define the orthogonal fields below:
\begin{eqnarray}
Z_\mu = \frac{1}{\sqrt{g^2+g^{\prime 2}}}(g X_\mu - g^\prime W_\mu ^z) ~,\label{zmi}\\
A_\mu = \frac{1}{\sqrt{g^2+g^{\prime 2}}}(g^\prime X_\mu + g W_\mu ^z) ~,\label{ami}\\
W_\mu^\dagger = W_\mu^x-i W_\mu ^y\label{wdag}~.
\end{eqnarray}
Using the above definitions we can put the matrix $M_\mu$ into the desired form
\[M_\mu =\left(\begin{array}{cc}  \frac{g}{\sin(\theta _w)}( \cos(2\theta _w) Z_\mu -\frac{1}{2}\sin(2 \theta _w) A_\mu) & 
- g^\prime W_\mu^\dagger \gamma ^0\\  g^\prime W_\mu \gamma ^0 &  \frac{g}{\sin(\theta _w)}  Z_\mu     \end{array}\right), \]
where the parameter $\theta _w$ is given by
\be
\label{wangle}
\sin(\theta _w) = \frac{g}{\sqrt{g^2+g^{\prime 2}}}~
\ee
and is known as the ``Weinberg angle".
Making use of the matrix $M_\mu$, (\ref{eqmass1}) and (\ref{eqmass2}) we write explicitly the 
full Lagrangian density in terms of the Dirac 4-component spinors:
\begin{eqnarray}
{\cal L}=i \bar{\psi _1} \gamma^\mu \partial _\mu \psi _1 + i \bar{\psi _2} \gamma^\mu \partial _\mu \psi _2 -m \bar{\psi} _1  \psi _1 \nonumber+\\
\frac{g}{\sin(\theta _w)} \bar{\psi}_1 \gamma ^\mu\left(\cos(2\theta _w) Z_\mu -\frac{1}{2}\sin(2 \theta _w) A_\mu\right) \psi _1\nonumber\\
-\frac{g}{\sin(\theta _w)} \bar{\psi}_2 \gamma ^\mu Z_\mu  \psi _2\nonumber\\
-\frac{g}{\tan(\theta _w)}(\bar{\psi}_1 \gamma ^\mu W_\mu^\dagger  
\gamma _0 \psi _2- \bar{\psi}_2 \gamma ^\mu W_\mu  \gamma _0 \psi _1)~.\label{LagDensity4}
\end{eqnarray}
In the above expression we omitted the hidden time coordinates and 
the indices are $\mu=(03,1,2,3)$. The reader can compare the resulting
lagrangian density with that obtained in the WSG electroweak theory.
At this point we emphasize that the theory being
developed here serves as a toy model to go one step further 
towards a gauge theory of electroweak interactions 
without the need of introducing a scalar Higgs field. In order to simplify
things we have not included chirality, which is an essential ingredient
in the real world. Therefore, our ``electroweak theory" is a theory without 
chirality and not totally realistic. However, we are able to produce
a theory in which the electron and neutrino fields naturally arise as 
parts of a single entity, the spinor $\psi$, and also the electron can
pick up mass while the neutrino remain massless without appealing to the
introduction of the scalar Higgs field. 
Looking at (\ref{LagDensity4}), we must identify the field $\psi _1$ with the electron and $\psi _2$ with the neutrino. 
Also the field $A_\mu$ is the photon field because it couples only to the electrons,
while the fields $Z_\mu$ and $W_\mu$ couples to the electrons and neutrinos as well.
We must add to the above lagrangian density a term for the gauge fields. The straightforward procedure
is to define physical fields by the commutator of the covariant derivatives:
we introduce the fields
\[G_{\mu\nu} =  [D_\mu,D_\nu]~.\]
The lagrangian of the gauge fields becomes\cite{[ryder]}:
\begin{eqnarray}
{\cal L}_{GF} = -\frac{1}{4}(\partial _\mu X_\nu -  \partial _\nu X_\mu)^2\nonumber\\
-\frac{1}{4}(\partial _\mu {\bf W}_\nu -  \partial _\nu {\bf W}_\mu + g^\prime {\bf W}_\mu \times {\bf W}_\nu)^2 \label{GfLag}
\end{eqnarray}
with indices $\mu,\nu = (01,02,03,1,2,3)$.
It is possible to keep the photon field massless and to give masses to the other gauge
fields using the same arguments that lead us to make the electron field massive while the neutrino
remained massless.  

In summary, we have constructed an electroweak theory without chirality, gauging
the rotational symmetry of the time degrees of freedom. The six dimensional
rotation group $SO(3,3)$ locally breaks into an $SO(1,3)$ Lorentz group, with just one 
time coordinate. There are many ways to embed the $SO(1,3)$ into the larger
group $SO(3,3)$. Two observers at infinitesimally separated points of spacetime may choose
a slightly different axis for the observable third time coordinate. When passing from one
point to another of the spacetime this distinction of the time axis are not directly 
observable and both observers may say that the symmetry group is $SO(1,3)$. This
gauge freedom in the choice of the observed time axis materializes as the existence of 
gauge fields. We may interpret the gauge fields as a manifestation of the
existence of extra time dimensions that are locally hidden. One next step 
in our theory is to introduce a chirality symmetry transformation, which discriminates left and right-handed
particles. To the best of our current knowledge such symmetry 
is still a mistery and needs to be investigated
further. \\

{\bf Acknowledgements}

C.A. Dartora would like to thank CNPq (Conselho Nacional de Pesquisa e Desenvolvimento) 
for partial financial support through
the project $555517/2006-3$ (Edital MCT/CNPq  42/2006).\\

\end{document}